\documentclass[aps,prb,twocolumn,superscriptaddress,longbibliography,showpacs]{revtex4-2}
\usepackage{amsmath,amssymb,wasysym,graphicx}
\usepackage[utf8]{inputenc}
\usepackage[T1]{fontenc}
\usepackage{xcolor}
\usepackage{ulem}

\IfFileExists{newtxtext.sty}
{\usepackage{newtxtext,newtxmath}}
{\IfFileExists{stix.sty}
	{\usepackage{stix}}
	{\IfFileExists{mathptmx.sty}
		{\usepackage{mathptmx}}{} } }

\usepackage{textcomp}

\usepackage{bm}

\IfFileExists{siunitx.sty}{\usepackage{booktabs,siunitx}}{}

\usepackage{color}
\definecolor{LinkColor}{rgb}{0.256,0.439,0.588}
\usepackage{hyperref}
\hypersetup{
	pdfauthor={good guys},
	pdftitle={good title},
	colorlinks=true,
	citecolor=LinkColor,
	linkcolor=LinkColor,
	urlcolor=LinkColor
}

\newcommand{\beq} {\begin{equation}}
\newcommand{\eeq} {\end{equation}}
\newcommand{\bea} {\begin{eqnarray}}
\newcommand{\eea} {\end{eqnarray}}
\newcommand{\be} {\begin{equation}}
\newcommand{\ee} {\end{equation}}

\newcommand{\ket}[1]{\left|#1\right>}
\newcommand{\bra}[1]{\left<#1\right|}

\begin{document}

\title{Quantum phase transition and critical behavior between the gapless topological phases}

\author{Hao-Long Zhang}
\altaffiliation{The first two authors contributed equally.}
\affiliation{Fujian Key Laboratory of Quantum Information and Quantum Optics, College
    of Physics and Information Engineering, Fuzhou University, Fuzhou, Fujian 350108, China}
    
\author{Han-Ze Li}
\altaffiliation{The first two authors contributed equally.}
\affiliation{Institute for Quantum Science and Technology, Shanghai University, Shanghai 200444, China}
\affiliation{Department of Physics, Shanghai University, Shanghai 200444, China}
\affiliation{School of Physics and Optoelectronics, Xiangtan University, Xiangtan 411105, China}

\author{Sheng Yang}
\email{qamber_ys@zju.edu.cn}
\affiliation{Institute for Advanced Study in Physics and School of Physics, Zhejiang University, Hangzhou 310058, China}

\author{Xue-Jia Yu}
\email{xuejiayu@fzu.edu.cn}
\affiliation{Fujian Key Laboratory of Quantum Information and Quantum Optics, College
    of Physics and Information Engineering, Fuzhou University, Fuzhou, Fujian 350108, China}

%-------------------------------------
\begin{abstract}
The phase transition between gapped topological phases represents a class of unconventional criticality beyond the Landau paradigm. However, recent research has shifted attention to topological phases without a bulk gap, where the phase transitions between them are still elusive. In this work, based on large-scale density matrix renormalization group techniques, we investigate the critical behaviors of the extended quantum XXZ model obtained by the Kennedy-Tasaki transformation. Using fidelity susceptibility as a diagnostic, we obtain a complete phase diagram, which includes both topological nontrivial and trivial gapless phases. Furthermore, as the XXZ-type anisotropy parameter $\Delta$ varies, both the critical points $h_c$ and correlation length exponent $\nu$ remain the same as in the $\Delta=0$ case, characterized by $c=3/2$ (Ising + free boson) conformal field theory. Our results indicate that fidelity susceptibility can effectively detect and reveal a stable unconventional critical line between the topologically distinct gapless phases for general $\Delta$. This work serves as a valuable reference for further research on phase transitions within the gapless topological phase of matter.

\end{abstract}

%-------------------------------------
\date{\today}
\maketitle

%%%%% Main Text %%%%%%%%%
%-------------------------------------
%-------------------------------------
\section{Introduction} 
\label{sec:introduction}

The classification of the quantum phase of matter constitutes a core issue in condensed matter and statistical physics~\cite{sachdev_2011,sachdev2023quantum,fradkin2013field,cardy1996scaling}. Nevertheless, in the past few decades, the development of topological phases has received significant attention~\cite{hasan2010rmp,qi2011rmp,gu2009prb,wen2017rmp,senthil2015symmetry,yu2024prl}, expanding our understanding of quantum phases beyond the Landau paradigm. A notable example is the symmetry-protected topological (SPT) phases~\cite{gu2009prb,wen2017rmp,senthil2015symmetry}, where the bulk is gapped and nontrivial gapless modes emerge at the boundary. It's worth emphasizing that discussions of SPT phases typically focus on gapped quantum phases~\cite{chen2014symmetry,wen1990to,chen2011prb,chen2010prb,chen2011prb_b,wen2014prb,chen2013prb,pollmann2012prb,wen2012prb}. Despite the crucial role of the bulk gap in defining topological phases, recent research~\cite{keselman2015prb,meng2011prb,fidkowski2011prb,kestner2011prb,Iemini2015prl,Lang2015prb,scaffidi2017prx,ruhman2017prb,JIANG2018753,verresen2018prl,keselman2018prb,verresen2020topology,verresen2021prx,duque2021prb,thorngren2021prb,umberto2021sci_post,yu2022prl,parker2018prb,yu2024universal,yu2024quantum,zhong2024topological,li2023intrinsicallypurely,li2023decorated,huang2023topological,wen2023prb,wen2023classification,su2024gapless} has revealed that many key features of topological physics, such as degenerate edge modes, persist in the gapless systems, even in the presence of nontrivial coupling between the boundary and critical bulk fluctuations, which we refer to as gapless topological phases. These new phases are not described by the Landau paradigm, and recent studies have explored their exotic properties~\cite{yu2022prl,yu2024universal,yu2024quantum,zhong2024topological,li2023intrinsicallypurely,li2023decorated,huang2023topological,wen2023prb,wen2023classification,su2024gapless}. However, the phase transitions between them remain largely unexplored.

\begin{figure}
    \centering
    \includegraphics{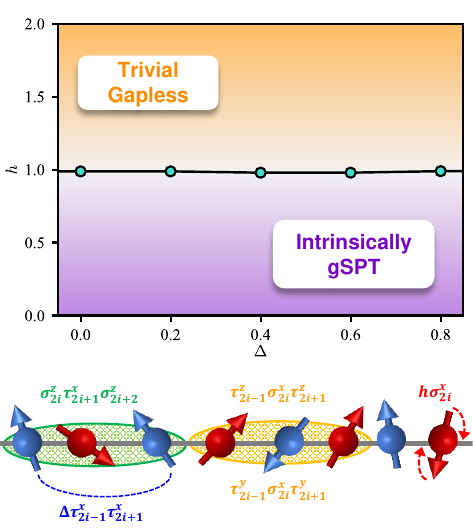}
    \caption{Phase diagram and schematic representation of the extended quantum XXZ spin chain with anisotropy parameter $\Delta$ and transverse field $h$. The critical point $h_{c}^{*}$ is obtained from polynomial fitting $h_c(N) = h_c^* + aN^{-1/\nu}$ of the peak position of fidelity susceptibility $h_c(N)$ for $N=48, 56, 64, 72, 80, 88, 96$ sites. Symbols denote the numerical results of the critical values $h^*_c$.}
    \label{fig1}
\end{figure}

On a different front, the development of the theory of quantum phase transitions (QPTs) stands as one of the striking achievements in modern physics~\cite{sachdev_2011,sondhi1997rmp,yu2022prb}. The traditional theory of phase transitions relies on the Landau-Ginzburg-Wilson symmetry-breaking paradigm~\cite{landau2013statistical}. However, in the last few decades, it has become clear that this paradigm does not fully capture the complexities at quantum critical points (QCPs), which are referred to as unconventional QCPs~\cite{xu2012unconventional,senthil2023deconfined,yang2023prb}. A notable example is the topological phase transitions between the gapped topological phases, which cannot be described in terms of fluctuating local order parameters or symmetry breaking~\cite{verresen2017prb,tsui2015quantum,tsui2017phase,lu2014prb}. Nonetheless, as mentioned in the last paragraph, the gapless topological phases represent a new type of exotic quantum matter beyond the Landau paradigm, and their phase transitions may be highly nontrivial and worth exploring in depth.

Fidelity susceptibility is a concept borrowed from quantum information theory and has found widespread utility as a useful diagnostic for pinpointing QCPs in the realms of condensed matter and statistical physics~\cite{albuquerque2010prb,yu2014fidelity,schwandt2009prl,yu2009pre,sun2015prb,konig2016prb,wei2019pra,yu2022prb_b, Tu_2023,zhang2023critical,yu2023pra,yu2023pre,guo2022pra}. Its advantage lies in the fact that it does not require prior knowledge of order parameters or symmetry breaking. To date, fidelity susceptibility has proven effective in detecting various types of QCPs, including conventional symmetry-breaking QCPs~\cite{albuquerque2010prb,schwandt2009prl}, topological phase transitions~\cite{sun2015prb}, Anderson transitions~\cite{wei2019pra}, non-conformal commensurate-incommensurate transitions~\cite{yu2022prb_b}, deconfined quantum criticality~\cite{sun2019prb}, and even non-Hermitian critical points~\cite{Tu_2023,sun2022biorthogonal}. Nevertheless, it remains an open question whether fidelity susceptibility can detect quantum phase transitions between gapless topological phases, and more importantly, determine the critical exponents and universality class at these QCPs.

In this work, we take the first step towards addressing the above questions by investigating the QPT between trivial and intrinsically gapless topological phases (gSPT). We accomplish this by constructing a one-dimensional extended quantum XXZ spin model through the Kennedy-Tasaki (KT) transformation~\cite{li2023intrinsicallypurely}. Specifically, using fidelity susceptibility as a diagnostic and in combination with the string order parameter and entanglement spectrum, we establish a complete global phase diagram, which includes both intrinsically gSPT and trivial gapless phases. Furthermore, by performing finite-size scaling on fidelity susceptibility, we conclude that as XXZ-type anisotropy parameters $\Delta$ vary, these topologically distinct gapless phases undergo a continuous phase transition, with the critical points $h_c$ and correlation length exponent $\nu$ remaining the same as in the $\Delta=0$ case, characterized by conformal field theory (CFT) with central charge $c=3/2$~\cite{francesco2012conformal,ginsparg1988applied}, which can be identified as an Ising CFT combined with a free boson CFT~\cite{li2023intrinsicallypurely}. It indicates that the unconventional critical point between topologically distinct gapless phases for $\Delta=0$ expands to a critical line for general $\Delta$.

The paper is organized as follows: Sec.~\ref{sec:model} contains the lattice model of the extended quantum XXZ spin chain after the KT transformation, the numerical method employed, the string order parameter, entanglement spectrum, and the scaling relations of fidelity susceptibility. Section ~\ref{sec:results} shows the global phase diagram of the model and the finite-size scaling of the various physical quantities. The conclusion is presented in Sec.~\ref{sec:con}. Additional data for our numerical calculations are provided in the Appendix.

\section{Model and Method}
\label{sec:model}
\subsection{Extended quantum XXZ spin chain through KT transformation}

We consider a lattice model exhibiting topologically distinct gapless quantum phases. This model can be obtained by stacking an Ising-ordered Hamiltonian with an XXZ chain via the KT transformation~\cite{li2023intrinsicallypurely}. The Hamiltonian is given by~\cite{yu2024universal}:

\begin{equation}
\label{HigSPT}
\begin{split}
  H = - \sum_{i=1}^{L} \Big( & \tau_{2i-1}^{z}\sigma_{2i}^{x}\tau_{2i+1}^{z} + \tau_{2i-1}^{y}\sigma_{2i}^{x}\tau_{2i+1}^{y} + \Delta\tau_{2i-1}^{x}\tau_{2i+1}^{x} \\
&  + \sigma_{2i}^{z}\tau_{2i+1}^{x}\sigma_{2i+2}^{z} + h\sigma_{2i}^{x} \Big), 
\end{split}
\end{equation} 
where $L$ denotes the number of unit cells, with the total number of sites $N$ being twice that, i.e., $N=2L$. Each unit cell is composed of a pair of spins $(\tau_{2i-1}, \sigma_{2i})$ represented by Pauli operators $\sigma^{\alpha}$ and $\tau^{\alpha}$ on the even and odd sites, respectively. The parameter $h > 0$ acts only on even sites ($\sigma$ spins) and denotes the strength of the transverse field. The XXZ-type anisotropic parameter $\Delta$ renders the Hamiltonian non-integrable and exact solvability is not feasible. The system exhibits a class of gapless phases described by a $c=1$ free boson CFT and possesses a $\mathbb{Z}_{4}$ symmetry generated by $U = \prod_{i}\sigma^{x}_{2i}e^{i\frac{\pi}{4}(1-\tau^{x}_{2i-1})}$, and also exhibits an emergent anomaly in the low-energy sector, known as intrinsically gSPT phases~\cite{thorngren2021prb}. Specifically, in this sector, the $\mathbb{Z}_{4}$ symmetry is approximately realized as $U \sim \prod_{i}\sigma^{x}_{2i}e^{i\frac{\pi}{4}(1-\sigma^z_{2i-2} \sigma^z_{2i})}$, which is analogous to the anomaly observed on the boundary of a 2+1D Levin-Gu SPT state~\cite{levin2012prb}. This anomaly prevents the system from realizing a unique symmetry-preserving gapped phase and gives rise to interesting physical properties. Furthermore, in an open chain with a length $N$, the square of the low-energy symmetry operator fractionalizes onto each end of the boundary, as detailed in~\cite{wen2023classification,wen2023prb}. Specifically, $U^2 \sim \tau^x_1 \sigma_2^z \sigma^z_{2L}$. This charge locally anticommutes with the $U$ symmetry, protecting a two-fold degeneracy in the intrinsically gSPT ground state.

Since the model is not integrable and exact solvability is not feasible, in this work, we solve the model using the density matrix renormalization group (DMRG) method~\cite{white1992prl,white1993prb,SCHOLLWOCK201196,schollwock2005rmp} based on matrix product states (MPS)\cite{vidal2003prl,vidal2003prl_b,vidal2004prl}. DMRG stands as one of the most powerful unbiased numerical techniques for addressing one-dimensional strongly correlated many-body systems. We have fixed the maximal MPS bond dimension at $1024$ to ensure reliable convergence of true energy eigenstates and fidelity susceptibilities. To this end, we maintain relative energy errors below $10^{-9}$. The fidelity susceptibility, defined later [Eq.(\ref{chiF})], is computed with a minimal step of $\delta h=10^{-3}$. In practical DMRG calculations, a random initial state is chosen, and open boundary conditions are applied in most cases.

\subsection{String order and entanglement spectrum in gapless topological phases}
We first utilize the long-distance behavior of non-local string order parameters and the bulk entanglement spectrum under periodic boundary conditions (PBC) to identify the possible quantum phases in the phase diagram before investigating QPT and pinpointing QCPs.

Following the KT transformation, as described in Ref.~\cite{li2023intrinsicallypurely}, the conventional local spin correlation function before the KT transformation is converted into string order parameters afterward, specifically denoted as $\sigma$ and $\tau$ string correlations:

\begin{align}
    \mathcal{O}_{\sigma}(\vert{i-j}\vert) & = \langle{\sigma_{2i}^{z}(\prod_{k=i}^{j-1}\tau_{2k+1}^{x})\sigma_{2j}^{z}}\rangle \,, \\
    \mathcal{O}_{\tau}(\vert{i-j}\vert) & = \langle{\tau_{2i-1}^{z}(\prod_{k=i}^{j-1}\sigma_{2k}^{x})\tau_{2j-1}^{z}}\rangle \,.
\end{align}
For $\Delta=0.0$, in the trivial gapless phase ($1.0<h<2.0$), the $\sigma$ string correlation exhibits exponential decay at long distances, while the $\tau$ string correlation function displays algebraic decay behavior. Conversely, in the intrinsically gSPT phase ($0.0<h<1.0$), the $\sigma$ string correlation exhibits long-range order, while the $\tau$ string correlation still displays algebraic decay behavior.

Furthermore, the bulk entanglement spectrum encodes information beyond entanglement entropy, suggesting that the bulk ground state wave function captures universal boundary information, such as topologically protected degenerate edge modes~\cite{pollmann2010prb}, and hence can be used to detect the gapped/gapless topological phases~\cite{zache2022entanglement,yu2024universal}. The entanglement spectrum consists of the eigenvalues of the entanglement Hamiltonian $\Tilde{H}_{A}$, related to the reduced density matrix ($\rho_{A}$) of subsystem $A$ by:
\begin{equation}
     \label{E_rho}
     \rho_{A} = \text{Tr}_{B}\ket{\Psi}\!\bra{\Psi} = \sum_{\alpha} e^{-\xi_{\alpha}} \ket{\Psi_{\alpha}^{A}}\bra{\Psi_{\alpha}^{A}} = e^{-\Tilde{H}_{A}}\,.
\end{equation}
Here, $\ket{\Psi}$ represents the ground state wave function of the Hamiltonian and $\xi_{\alpha} \equiv -\ln\lambda_{\alpha}$ where $\lambda_{\alpha}$ is the eigenvalue of $\rho_{A}$. In our study of 1D quantum chains, $A={1,2,...L/2}$ and $B={L/2+1,...L}$ represent a spatial bipartition of the entire chain, and the bulk entanglement spectrum displays two degenerate and non-degenerate ground states in topologically nontrivial and trivial gapless phases, respectively.

\begin{figure}
    \centering
    \includegraphics[width=0.95\columnwidth]{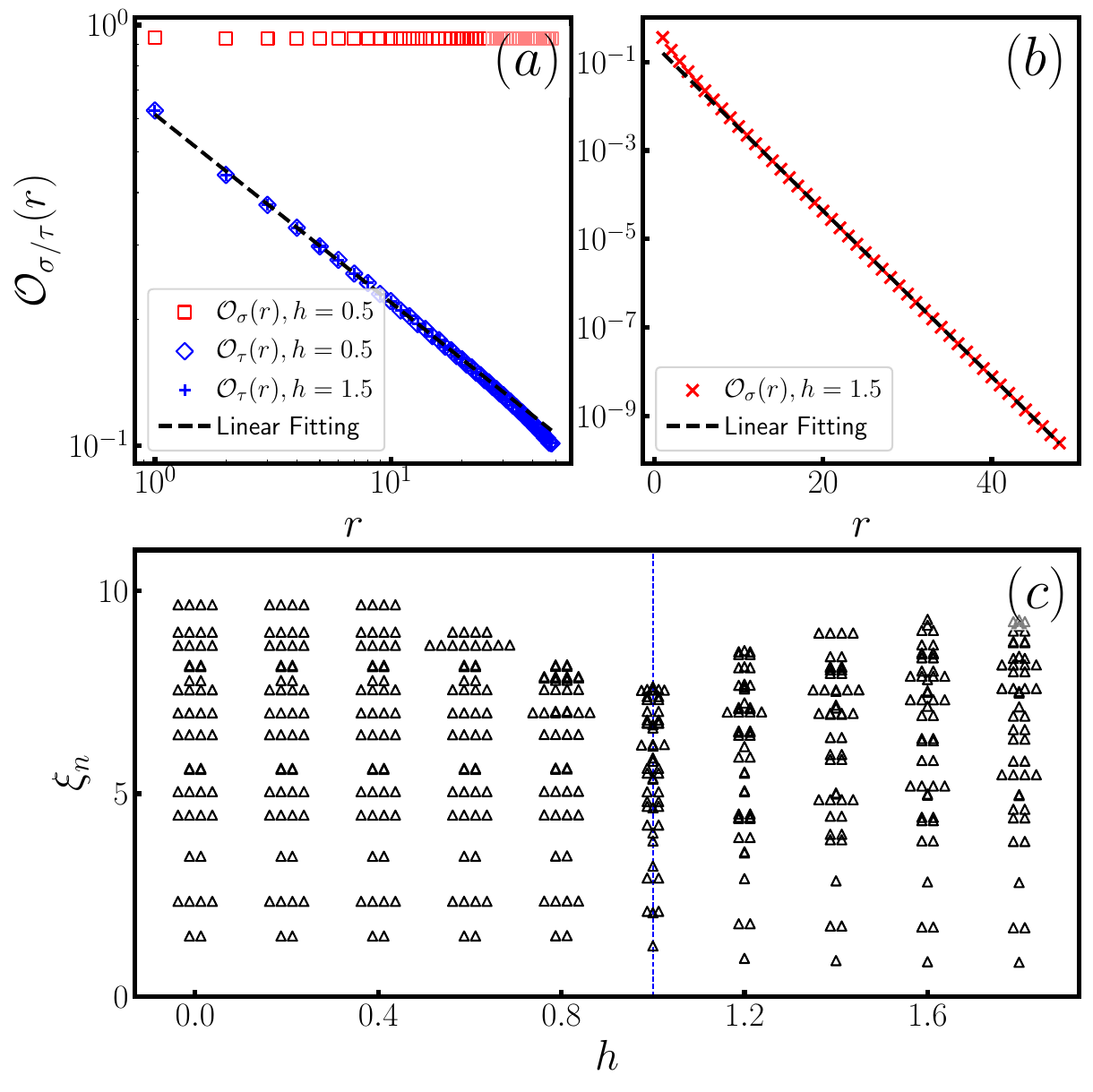}
    \caption{(a-b) The scaling behaviors of $\tau$ and $\sigma$ string correlations in the intrinsically gSPT phase ($h=0.5$) and trivial gapless phase ($h=1.5$) for $\Delta=0.2$ and $N=192$. (c) The evolution of the entanglement spectrum as a function of $h$ for $\Delta=0.2$ and $N=96$ under PBC with only the first $50$ low-lying values are displayed in the plot.}
    \label{sop_es}
\end{figure}

\subsection{Fidelity susceptibility and scaling relations}
In this work, we utilize fidelity susceptibility, a concept borrowed from quantum information theory, offering a remarkably simple and intuitive method for identifying QCPs and obtaining critical exponents through finite-size scaling.

The concept of fidelity susceptibility is as follows: Given a Hamiltonian $H(h) = H_0 + h H_1$ with a driving parameter $h$, the quantum ground-state fidelity $F(h,h+\delta h)$ is defined as the overlap amplitude of two ground states $\ket{\psi(h)}$ and $\ket{\psi(h + \delta h)}$:
\begin{equation}
    F(h,h+\delta h) = |\langle \psi(h)|\psi(h+\delta h)\rangle|.
\end{equation}
When a system undergoes a continuous phase transition from an ordered to a disordered phase by tuning the external field $h$ to a critical value $h_{c}^{*}$, at which the structure of the ground-state wave function changes significantly, the quantum ground-state fidelity is nearly zero near $h_{c}^{*}$.  In practice, a more convenient quantity for characterizing QPTs is the fidelity susceptibility, defined by the leading term of fidelity:
\begin{equation} \label{chiF}
    \chi_F(h) = \lim_{\delta h \rightarrow 0}\frac{2(1-F(h,h+\delta h))}{(\delta h)^2}.
\end{equation}

For a continuous QPT of a finite system with size $N$, fidelity susceptibility exhibits a peak at the pseudocritical point $h_c(N)$, and the true QCP $h_c^*$ can be estimated through polynomial fitting $h_c(N)=h_c^* + aN^{-1/\nu}$~\cite{sandvik2010computational}. In the vicinity of $h_c(N)$, previous studies have shown that the finite-size scaling behaviors of fidelity susceptibility $\chi_F(h)$ are described by~\cite{albuquerque2010prb}
\begin{equation}\label{chiF_mu}
    \chi_F(h\rightarrow h_c(N)) \propto N^{\mu},
\end{equation}
where $\mu=2+2z-2\Delta_V$ is the critical adiabatic dimension. Here, $z$ is the dynamical exponent and $\Delta_V$ is the scaling dimension of the local interaction $V(x)$ at the critical point. On the other hand, it is shown that the fidelity susceptibility per site scales as~\cite{albuquerque2010prb}:
\begin{equation}\label{dc}
    N^{-d}\chi_F(h)  = N^{(2/\nu)-d}f_{\chi_F}(N^{1/\nu}\vert h-h_c^* \vert ),
\end{equation}
where $d$ is the spatial dimension of the system, $f_{\chi_F}$ is an unknown scaling function and $\nu$ is the critical exponent of the correlation length, which can be easily computed according to the relation $\nu=2/\mu$. Based on Eq.~(\ref{chiF_mu}) and (\ref{dc}), the values of critical exponent $\mu$ and $\nu$ can be determined and the corresponding critical behavior can be easily confirmed. In practice, the critical exponent $\nu$ and $\mu$ are usually extracted from fidelity susceptibility per site, $\chi_N(h) = \chi_F(h)/N^d$.
\begin{figure}
    \centering
    \includegraphics[width=0.95\columnwidth]{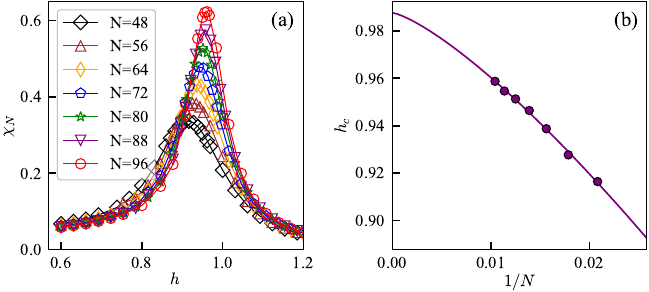}
    \caption{(a) Fidelity susceptibility per site $\chi_F/N$ of the extended quantum XXZ spin chain for $\Delta=0.2$ and $N=48, 56, 64, 72, 80, 88, 96$ sites as a function of the transverse field $h$; symbols denote DMRG results. (b) The extrapolation of critical point $h_c^*$ with different $N$; symbols denote the finite-size DMRG results for $\Delta=0.2$ and $N=48, 56, 64, 72, 80, 88, 96$ sites. We use polynomial fitting $h_c(N) = h_c^* + aN^{-1/\nu}$ and extrapolate the critical point $h_c^*\approx0.98761(8)$.}
    \label{fig2}
\end{figure}

\section{PHASE DIAGRAM AND CRITICAL BEHAVIOR}
\label{sec:results}
\subsection{Quantum phase diagram: an overview}
Before exploring phase transitions, let’s investigate the possible quantum phases that appear in a phase diagram. As a first step, we examine a limiting case: When $\Delta=0.0$, the model simplifies to an Ising-ordered Hamiltonian combined with an XY chain via the KT transformation. References~\cite{li2023intrinsicallypurely,li2023decorated} indicate that this model exhibits a phase transition between the intrinsically ($0.0<h<1.0$) and trivial gapless topological phases ($h>1.0$). For a general $\Delta$, since the model is no longer exactly solvable, we employ the DMRG simulations to ascertain the possible phases in the model by computing the scaling behavior of $\sigma$ and $\tau$ string correlations for $\Delta=0.2$ (see Appendix~\ref{sec:SM1} for other $\Delta$). As depicted in Fig~\ref{sop_es}(a) and (b), we observed that when $h<1.0$, $\sigma$ and $\tau$ string correlations respectively exhibit long-range order and power-law decay behavior at the long-distance limit, consistent with the characteristics of intrinsically gSPT as we mentioned before. Conversely, when $h>1.0$, $\sigma$ and $\tau$ string correlations display exponential and power-law decay behaviors, respectively, consistent with the features of trivial gapless phases.

Furthermore, to elucidate the topological properties of gapless topological phases more clearly, we computed the bulk entanglement spectrum as a function of $h$ under $\Delta=0.2$ (see Appendix~\ref{sec:SM1} for other $\Delta$). As illustrated in Fig~\ref{sop_es}(c), we observe that the ground state degeneracy of the bulk entanglement spectrum transforms from twofold ($0.0<h<1.0$) to a unique ground state ($h>1.0$). This indicates that the system undergoes a phase transition from topologically nontrivial to trivial gapless phases.

The numerical results mentioned above suggest that irrespective of the magnitude of the XXZ-type perturbation, which breaks the integrability of the model, stable trivial and intrinsically gapless topological phases continue to exist in the ground state phase diagram, determined through DMRG calculations for $N=48, 56, 64, 72, 80, 88, 96$ sites, with the results presented in Fig.~\ref{fig1}. When $\Delta=0.0$, the ground state exhibits an intrinsically gSPT phase with a two-fold degeneracy at $h\in (0.0,  1.0)$ and transitions to a trivial gapless phase as $h$ greater than $1.0$, consistent with previous findings\cite{li2023intrinsicallypurely}. Additionally, for finite $\Delta$, we find that the model exhibits a stable intrinsically gSPT and a trivial gapless phase across the entire range of $\Delta$ that we consider.

\begin{figure}
    \centering
    \includegraphics[width=1.0\linewidth]{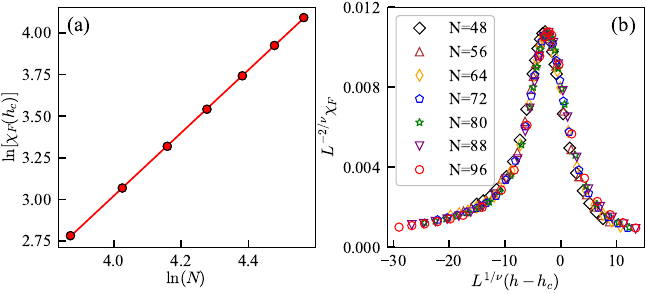}
    \caption{(a) The finite-size scaling of the fidelity susceptibility per site $\chi_F(h_c)$ at the peak position $h_c$ where $\mu\approx1.894$. (b) Data collapse of fidelity susceptibility $\chi_F$ for the extended quantum XXZ spin chain; symbols denote the finite-size DMRG results for $\Delta=0.2$ and $N=48, 56, 64, 72, 80, 88, 96$ sites, where $\nu\approx 1.05873(1)$, $\mu\approx 1.88905(3)$ and $h_c^*\approx 0.98761(8)$ are used for data collapse plots.}
    \label{fig3}
\end{figure}

The finite-size scaling behavior of fidelity susceptibility for $\Delta=0.2$ with different $N$ is presented in Fig.~\ref{fig2}(a), which follows the scaling relation $\chi_{N}(h_{c}(N)) \propto N^{\mu-1}$ [Eq.~(\ref{chiF_mu})] near the second-order QPT critical point. As the system size $N$ increases, the peak position $h_{c}(N)$ approaches the exact critical point value $h_c^*$ more closely (see Appendix~\ref{sec:SM2} for other $\Delta$). Specifically, for the extended XXZ model with $\Delta=0.2$, $h_c^*$ is determined by polynomial fitting $h_c(N)=h_c^* + aN^{-1/\nu}$, and then extrapolating to $N \to \infty$ [Fig.~\ref{fig2}(b)]. According to Eq.~(\ref{dc}), the fidelity susceptibility follows an exact scaling relation and collapses to one master curve [Fig.~\ref{fig3}(b)], confirming the appropriateness of the extrapolation. The finite-size scaling behavior of fidelity susceptibility for other $\Delta$ values is also investigated (see Appendix~\ref{sec:SM3} for details), and the results are presented in Table~\ref{tab1}. The findings indicate that the QCPs remain unchanged as $\Delta$ varies.

\begin{table}[b]
\centering
\caption{\label{tab1}Critical exponents and critical points of the extended quantum XXZ chain for different $\Delta$.}
\begin{ruledtabular}
\begin{tabular}{cccc}
  $\Delta$ & $h_c^*$ & $\mu$ & $\nu$\\
\colrule
0.0& 0.98761(8)& 1.88965(3)& 1.05839(1)\\
0.2& 0.98761(8)& 1.88905(3)& 1.05873(1)\\
0.4& 0.97875(1)& 1.88995(3)& 1.05822(1)\\
0.6& 0.97875(1)&  1.88925(3)& 1.05862(1)\\
0.8& 0.98903(9)& 1.88958(2)& 1.05843(1)\\
\end{tabular}
\end{ruledtabular}
\end{table}

\subsection{Finite-size scaling and critical exponent} 
The next questions concern the critical behavior of the extended XXZ chains with different $\Delta$ values and whether there exists a critical threshold $\Delta_c$ at which the critical behavior changes. To address these questions, we conduct large-scale DMRG simulations for various $N$ in the region $0.0 \le \Delta \textless 1.0$ to extrapolate the critical exponents $\mu$ and $\nu$ through the finite-size scaling of fidelity susceptibility.

The fidelity susceptibility per site, $\chi_{N}=\chi_{F}/N$, at the peak position $h_{c}(N)$ for different $N$ at $\Delta=0.2$ is illustrated in Fig.~\ref{fig3}(a). The adiabatic critical dimension $\mu$ is well-fitted by a polynomial function of $\chi_{N}(h_{c}(N))=N^{\mu-1}(c+dL^{-1})$.

According to Eq.(\ref{dc}), the fidelity susceptibility can be scaled by $N^{-2/\nu}\chi_{F}$ as a function of $N^{1/\nu}(h-h_{c}^{*})$ in the vicinity of the QCP $h_{c}^{*}$. The correlation length exponent $\nu$ is then determined by $\nu=2/\mu$. By substituting the obtained critical point $h_{c}^{*}$ and critical exponent $\nu$ into Eq.(\ref{chiF_mu}), all fidelity susceptibilities for different $N$ collapse into a single curve (Fig.~\ref{fig3}(b)), indicating the accuracy of the estimated critical point and critical exponent (see Appendix.~\ref{sec:SM4} for other $\Delta$). It is worth noting that the peak in the data collapse is not precisely at $0$ due to the finite-size effect for $h(N)=h_{c}^{*}+aL^{-1/\nu} (a\ne 0)$.

\begin{figure}
    \flushleft
    \includegraphics[width=1.0\linewidth]{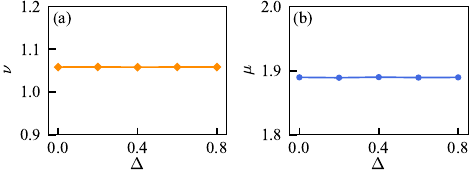}
    \caption{The correlation length exponent $\nu$ (a) and critical adiabatic dimension $\mu$ (b) with respect to $\Delta$; the symbols denote the finite-size DMRG results that are obtained by extrapolating from the fidelity susceptibility $\chi_{F}(h_{c}(N))$ at the peak position $h_{c}(N)$ of $N=48, 56, 64, 72, 80, 88, 96$ sites.}
    \label{fig4}
\end{figure}

The extrapolations of the critical adiabatic dimension $\mu$ and the correlation length exponent $\nu$ for other $\Delta$ values are presented in Appendix~\ref{sec:SM3}, and the results for all $\Delta$ are summarized in Table.~\ref{tab1} and Fig.~\ref{fig4}. Both $\nu$ and $\mu$ remain unchanged as $\Delta$ varies. This indicates that the XXZ-type term acts as irrelevant perturbations and the unconventional critical point described by CFT with central charge $c=3/2$ for $\Delta=0.0$, as discussed in the literature~\cite{li2023intrinsicallypurely}, expands to a critical line for general $\Delta$. This trend also suggests that the unconventional QCP between the topologically distinct gapless phases is robust against the XXZ-type perturbation, and there does not exist a critical threshold $\Delta_c$ at which the critical behavior changes.

\section{Conclusions and outlook}
\label{sec:con}
In summary, we investigate the phase transition between topologically distinct gapless phases, namely, intrinsically gSPT phases. We establish a complete phase diagram for the Hamiltonian, which is a one-dimensional extended XXZ model constructed by the KT transformation. Using fidelity susceptibility as a diagnostic and in combination with the string order parameter and entanglement spectrum, we unambiguously reveal the intrinsically gSPT and trivial gapless phases in the phase diagram. Moreover, by computing fidelity susceptibility and performing finite-size scaling, we observe a continuous phase transition between the topologically distinct gapless phases as the XXZ-type anisotropy term $\Delta$ varies. Remarkably, the critical points and correlation length exponent remain the same as in the $\Delta=0$ case, characterized by CFT with central charge $c=3/2$, which can be identified as an Ising CFT combined with a free boson CFT. Our results indicate that fidelity susceptibility can effectively detect and reveal an unconventional critical point for $\Delta = 0.0$ extended into a critical line for general $\Delta$. Future intriguing questions involve exploring the critical behavior between topologically distinct gapless phases in higher dimensions and within different symmetry groups (e.g., $\mathbb{Z}_{3}$, $U(1)$, among others), as well as constructing finite-temperature phase diagrams. Our work could shed new light on the phase transition between gapless topological phases of matter.

\begin{acknowledgments}
We thank Linhao Li for very helpful discussions. Numerical simulations were carried out with the ITENSOR package~\cite{itensor} on the Kirin No.2 High Performance Cluster supported by the Institute for Fusion Theory and Simulation (IFTS) at Zhejiang University. X.-J.Yu acknowledges support from the start-up grant No.511317 of Fuzhou University.

\end{acknowledgments}

\bibliography{main}

\onecolumngrid

\begin{appendix}

%\section*{Supplemental Material}

\renewcommand{\theequation}{A\arabic{equation}}
\setcounter{equation}{0}
\renewcommand{\thefigure}{A\arabic{figure}}
\setcounter{figure}{0}
\renewcommand{\thetable}{A\arabic{table}}
\setcounter{table}{0}

\section{STRING ORDER AND ENTANGLEMENT SPECTRUM FOR OTHER VALUES OF $\Delta$}
\label{sec:SM1}
In this section, we provide additional data to identify the intrinsically gSPT and trivial gapless phases through the scaling behavior of string correlations and the entanglement spectrum for different $\Delta$ values.

\begin{figure*}
    \centering
    \includegraphics[width=1.0\columnwidth]{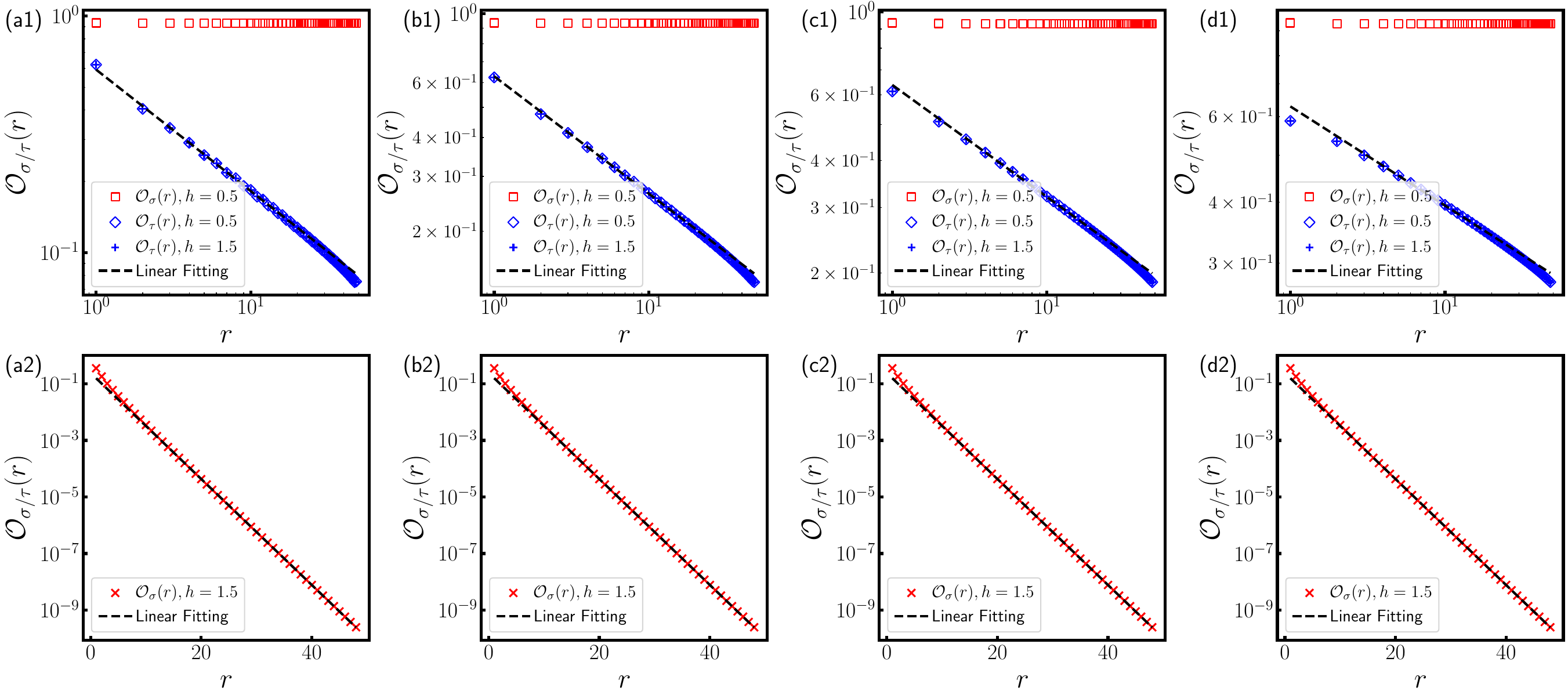}
    \caption{The scaling behaviors of string correlations in the intrinsically gSPT phase ($h=0.5$) and trivial gapless phase ($h=1.5$) for (a1-a2) $\Delta=0.0$, (b1-b2) $\Delta=0.4$, (c1-c2) $\Delta=0.6$, and (d1-d2) $\Delta=0.8$. The simulated system size is $N=192$ under OBC.}
    \label{correlations}
\end{figure*}

On the one hand, as described in the main text, the long-distance behavior of the $\tau$ and $\sigma$ string correlations can be completely different within the intrinsically gSPT and trivial gapless phases. Specifically, as shown in Fig.~\ref{correlations}, we computed the $\tau$ and $\sigma$ string correlations as a function of lattice distance $r$ for $\Delta = 0.0$ (a1-a2), $\Delta=0.4$ (b1-b2), $\Delta=0.6$ (c1-c2), and $\Delta=0.8$ (d1-d2) with a simulated system size of $N=192$ under OBC. The results indicate that $h>1$ and $h<1$ exhibit the characteristics of trivial gapless and intrinsically gapless SPT phases, respectively.

\begin{figure*}
    \centering
    \includegraphics[width=1.0\columnwidth]{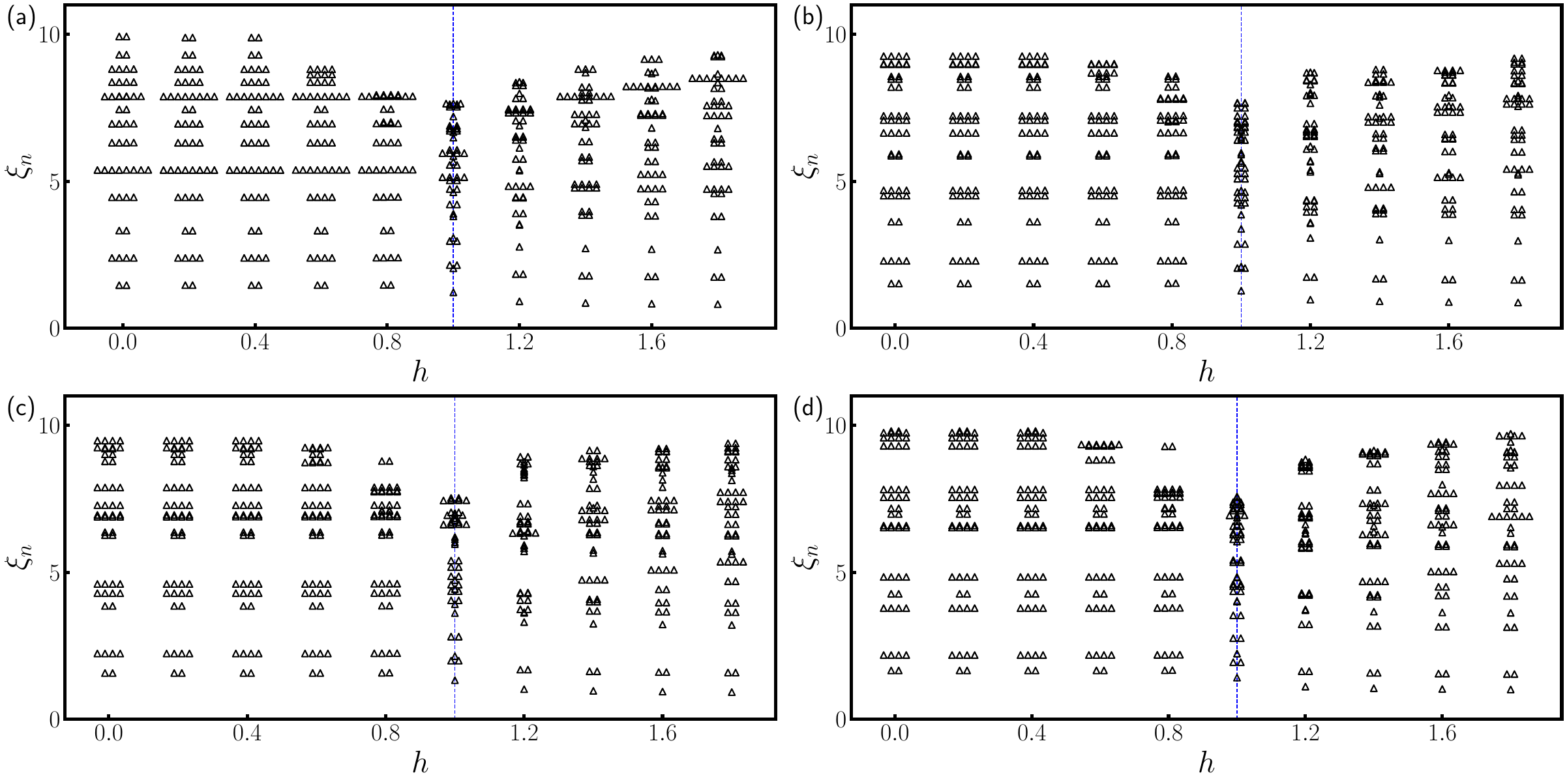}
    \caption{The entanglement spectrum as a function of $h$ for (a) $\Delta=0.0$, (b) $\Delta=0.4$, (c) $\Delta=0.6$, and (d) $\Delta=0.8$ with $N=96$ under PBC. We only display the first $50$ low-lying values in the plot.}
    \label{spectrum}
\end{figure*}

On the other hand, to more intuitively exhibit the topological properties of gapless quantum phases, as in the main text, we calculated the bulk entanglement spectrum as a function of $h$ under different $\Delta$, as shown in Fig.~\ref{spectrum} (a) $\Delta=0.0$, (b) $\Delta=0.4$, (c) $\Delta=0.6$, and (d) $\Delta=0.8$ with $N=96$ under PBC. Here, we only display the first $50$ low-lying values in the plot. Our results indicate that regardless of the magnitude of $\Delta$, the ground state degeneracy of the entanglement spectrum changes from double degeneracy to a unique ground state at $h \approx 1.0$ (blue dashed lines in Fig.~\ref{spectrum}) as $\Delta$ increases. This change in topological properties is consistent with the change in the long-distance behavior of string correlations as mentioned in the previous paragraph.

\section{FIDELITY SUSCEPTIBILITY FOR OTHER VALUES OF $\Delta$}
\label{sec:SM2}
In this section, we provide additional data to obtain the critical line via fidelity susceptibility for other values of $\Delta$.

\begin{figure}
    \centering
    \includegraphics[width=1.0\linewidth]{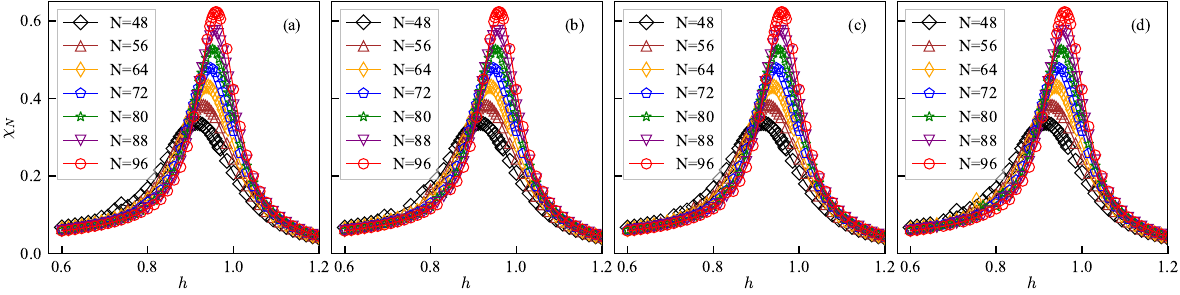}
    \caption{Fidelity susceptibility per site $\chi_N$ for the extended quantum XXZ spin chain for (a) $ \Delta =0.0$, (b) $\Delta = 0.4$, (c) $\Delta = 0.6$, (d) $\Delta = 0.8$ and $N=48, 56, 64, 72, 80, 88, 96$ sites as a function of driving parameter $h$; symbols denote finite-size DMRG results.}
    \label{fig5}
\end{figure}

As in the main text, fidelity susceptibility per site $\chi_N$ of the extended quantum XXZ spin chain for $\Delta = 0.0$ (a), $\Delta = 0.4$ (b), $\Delta = 0.6$ (c), and $\Delta = 0.8$ (d), with system sizes $N=48, 56, 64, 72, 80, 88, 96$, as a function of the transverse field $h$, are shown in Fig.~\ref{fig5}. At first, we observe that regardless of the value of $\Delta$, fidelity susceptibility exhibits obvious peaks as $h$ varies, indicating continuous phase transitions between topologically distinct gapless phases. Moreover, we find that the QCPs remain unchanged for different values of XXZ-type perturbation.

\section{QUANTUM CRITICAL POINT FITTING AND CRITICAL ADIABATIC DIMENSION FOR OTHER VALUES OF $\Delta$}
\label{sec:SM3}
In this section, we provide additional data to extrapolate accuracy critical points and critical adiabatic dimensions for other values of $\Delta$.

\begin{figure}
    \centering
    \includegraphics[width=1.0\linewidth]{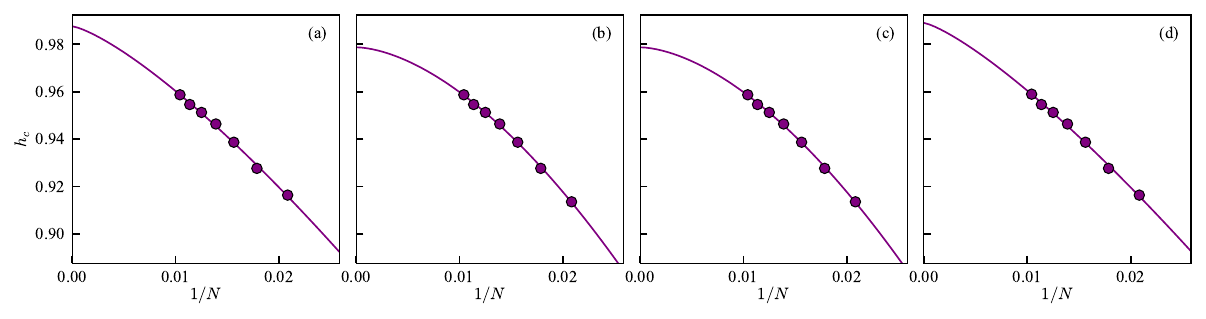}
    \caption{The finite-size scaling of pseudocritical point $h_c(N)$ as a function of inverse system size $1/N$ for (a)$ \Delta =0.0$, (b)$\Delta = 0.4$, (c)$\Delta = 0.6$, (d)$\Delta = 0.8$; We use polynomial fitting formula $h_c(N) = h_c^* + aN^{-1/\nu}$.}
    \label{fig8}
\end{figure}

As the same in the main text, on the one hand, we determined the pseudocritical point $h_{c}(N)$ corresponding to the maximum values of the fidelity susceptibility and performed finite-size scaling of the pseudocritical point $h_{c}(N)$ as a function of inverse system sizes $1/N$ for $\Delta=0.0$ (a), $\Delta=0.4$ (b), $\Delta=0.6$ (c), and $\Delta=0.8$ (d), with $N = 48, 56, 64, 72, 80, 88, 96$ sites, as shown in Fig.\ref{fig8}. The extrapolated critical points are summarized in Table~\ref{tab1}. The accurate critical point $h_{c}^{*}$ remains unchanged with increasing $\Delta$.

\begin{figure}
    \centering
    \includegraphics[width=1.0\linewidth]{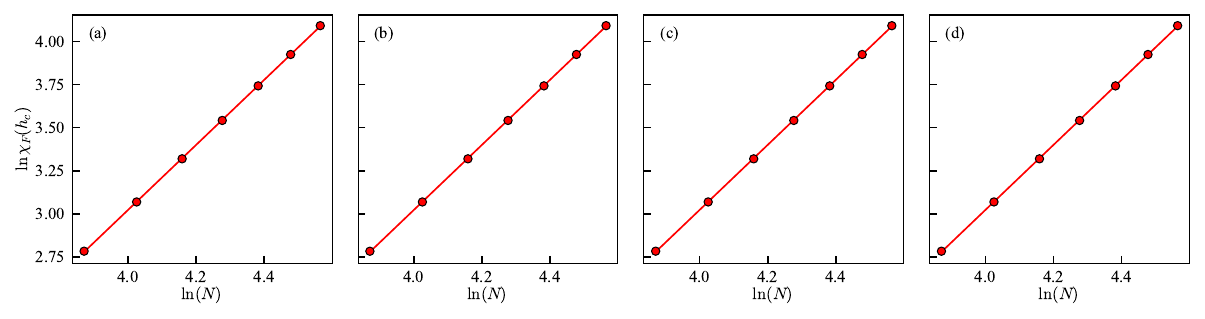}
    \caption{The maximal of fidelity susceptibility per site $\chi_F(h_c)$ as a function of system size $N$ for (a)$ \Delta =0.0$, (b)$\Delta = 0.4$, (c)$\Delta = 0.6$, (d)$\Delta = 0.8$, and $N=48, 56, 64, 72, 80, 88, 96$ sites. We use the fitting formula$\chi_{F}(h_c) = N^{\mu}(c+d N^{-1})$.}
    \label{fig7}
\end{figure}

Furthermore, we examined the maximal fidelity susceptibility per site $\chi_{N}(h_{c}(N)) = \chi_{F}(h_{c}(N))/N$ as a function of system sizes $N$ for $\Delta = 0.0$ (a), $\Delta=0.4$ (b), $\Delta=0.6$ (c), $\Delta=0.8$ (d), and $N=48,56,64,72,80,88,96$ sites, as illustrated in Fig.\ref{fig7}. The critical adiabatic dimensions are also summarized in Table~\ref{tab1}. We observe that the critical adiabatic dimension $\mu$ remains unchanged with increasing $\Delta$.

\section{DATA COLLAPSE FOR OTHER VALUES OF $\Delta$}
\label{sec:SM4}
In this section, we present additional data demonstrating the variation in correlation length exponents $\nu$ of the extended quantum XXZ chain.

\begin{figure}
    \centering
    \includegraphics[width=1.0\linewidth]{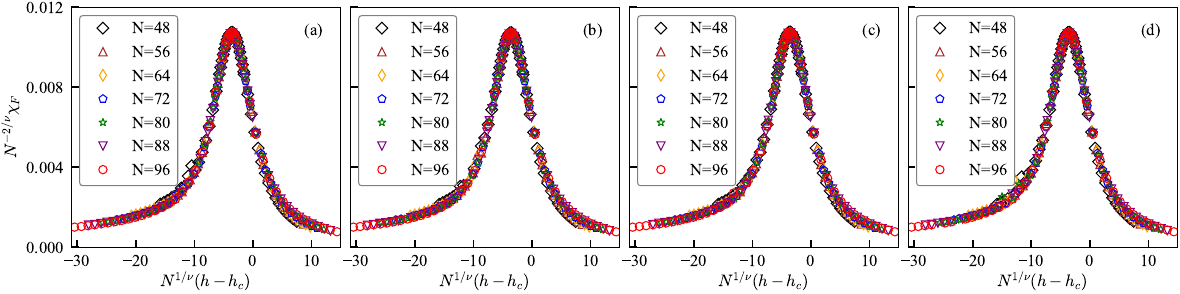}
    \caption{Data collapse of fidelity susceptibility $\chi_F$ for the extended quantum XXZ spin chain for (a) $ \Delta =0.0$, (b) $\Delta = 0.4$, (c) $\Delta = 0.6$, (d) $\Delta = 0.8$ and $N=48, 56, 64, 72, 80, 88, 96$ sites as a function of transverse field $h$.}
    \label{fig6}
\end{figure}

As the same in the main text, data collapse of fidelity susceptibility $\chi_{F}$ for the one-dimensional extended XXZ model is shown for $\Delta=0.0$ (a), $\Delta=0.4$ (b), $\Delta=0.6$ (c), and $\Delta=0.8$ (d), with $L = 48, 56, 64, 72, 80, 88, 96$ sites in Fig.~\ref{fig6}. The correlation length exponents are summarized in Table~\ref{tab1}. It is evident that the correlation length exponents of the extended quantum XXZ chain is the same as observed in the $\Delta=0.0$ case, characterized by a CFT with central charge $c=3/2$, which can be understood as a combination of Ising and free boson CFTs.

% \bibliography{SM}
\end{appendix}

\end{document}